\RequirePackage{fix-cm}
\documentclass[twocolumn,epjc3]{svjour3}  
\smartqed  
\RequirePackage{graphicx}
 \RequirePackage{mathptmx}      
\RequirePackage{latexsym}
\RequirePackage[numbers,sort&compress]{natbib}
\RequirePackage[colorlinks,citecolor=blue,urlcolor=blue,linkcolor=blue]{hyperref}
\RequirePackage{amsmath}
\journalname{Eur. Phys. J. A}
\begin{document}
\sloppy 
\title{Imprints of clustering in multiplicity fluctuations
}

\titlerunning{Imprints of clustering...}        

\author{A. Bazgir\thanksref{addr1}, 
V.Z. Reyna Ortiz\thanksref{addr1}, 
M. Rybczy\'{n}ski\thanksref{e1,addr1}, 
U. Shah\thanksref{addr1} 
        \and
        Z. W\l odarczyk\thanksref{addr1} 
}

\thankstext{e1}{e-mail: maciej.rybczynski@ujk.edu.pl}


\institute{Institute of Physics, Jan Kochanowski University, 25-406 Kielce, Poland \label{addr1}
}

\date{Received: date / Accepted: date}

\maketitle

\begin{abstract} In this paper, we investigate the multiplicity fluctuations of charged particles observed in high-energy nuclear collisions and relate them to the size of hadronizing systems which happen during such processes. We use the average multiplicities $\langle N\rangle $ and variances $Var\left(N\right)$ of multiplicity distributions of charged particles produced in centrality selected collisions of relativistic heavy-ion nuclei to evaluate the dynamic variance $\Omega$ and study its dependence on the size of colliding systems. We connect the observed system-size dependence of multiplicity fluctuations with the clustering phenomena and the finiteness of the hadronizing sources and the thermal bath.

\end{abstract}

\section{Introduction}
\label{sec:introduction}

One of the key methods for the study of strongly-interacting matter suffering extreme conditions is the measurement of event-by-event fluctuations of experimental observables. The registered fluctuations are sensitive to the proximity of the phase transition between hadronic gas state and the quark-gluon plasma (QGP), and the critical point of strongly interacting matter. Thus, they give us information on the dynamics hidden in the system formed in the collision~\cite{Stephanov1998, Stephanov1999, Koch:2001, Jeon2000, Jeon2003, Karsch:2005, Stokic:2008}. \\

Event-by-event fluctuations of number of particles produced in collisions of relativistic ions have been studied already at the CERN Super Proton Synchrotron (SPS) by the WA98~\cite{WA98}, NA49~\cite{NA49-1,NA49-2} and CERES~\cite{CERES-1} experiments, at the BNL Relativistic Heavy Ion Collider (RHIC) by the PHOBOS~\cite{PHOBOS} and PHENIX~\cite{PHENIX-1} experiments, and at the CERN Large Hadron Collider (LHC) by the ALICE experiment~\cite{ALICE-1}. Experimental measurements of physical fluctuation observables in collisions of ions can provide some important signals useful to investigate the response of a colliding system to external perturbations. Fluctuations in a finite system being in contact with finite thermal bath are discussed in the~\ref{sec:appendix}. Using standard theoretical developments it is possible to extract quantities related to the thermodynamic properties of the system. These quantities include entropy, chemical potential, viscosity, specific heat, and isothermal compressibility~\cite{VanHove:1983,Mrowczynski:1997}. \\

In particular, the ability of hadronic matter to generate some spatial structures (clusters) with self-similar multiplicity fluctuations indicates the self-organiziation ability of strongly-interacting matter~\cite{Mryb}. In the collisions of relativistic ions a hot QGP is produced, which then cools down and transits to a hadron gas. The so-called self-organized criticality is the appropriate mechanism leading to a universal scale-free behavior~\cite{Castorina4}. Self-organized criticality~\cite{Bak5} is a property of non-equilibrium dynamical systems that have a critical point as an attractor. The macroscopic properties of such systems are characterized by the spatial and/or temporal scale-invariance of the phase transition critical point. An interesting feature of strongly interacting matter is its tendency to self-organize. One striking instance of this ability to generate spatial structures is the cluster phase, where clusters broadly distributed in size constantly move and evolve through particle exchange~\cite{Castorina4}. \\

The multiplicity fluctuations registered on an event-by-event basis supply us with information on the mechanisms of particle production. Usually the magnitude of multiplicity fluctuations is quantified by scaled variance of multiplicity distribution:
\begin{equation}
\omega = \frac{Var(N)}{\langle N \rangle},
\label{eq:omega}
\end{equation}
where $Var\left(N\right)=\sum_{N}\left(N-\langle N\rangle\right)^{2}\cdot P\left(N\right)$ is the variance of the multiplicity distribution, $\langle N\rangle=\sum_{N} N\cdot P\left(N\right)$ is the average multiplicity, and $P\left(N\right)$ is the charged-particle multiplicity distribution. Scaled variance of multiplicity distribution scales with collision centrality in heavy ion collisions, when centrality of collision is expressed by number of nucleons participating in the collision, $N_{part}$. It is interesting to study the ratio of the variance of multiplicity distribution to the square of the average multiplicity. This quantity as a function of $N_{part}$ shows an intriguing power-law dependence:
\begin{equation}
\frac{Var(N)}{\langle N \rangle^2} \sim N_{part}^{-\alpha}
\label{eq:omega2}
\end{equation}
with the exponent $\alpha \simeq 1.25 $. This intriguing scaling first observed by the PHENIX Collaboration~\cite{PHENIX-2}, also holds for the ALICE data~\cite{ALICE-1}. 
When applied a description of the grand-canonical ensemble, we expect $\alpha=1$~\cite{Mrowczynski:1997}. In the following we attribute the observed system size dependence of the magnitude of fluctuations to the clustering phenomena.

\section{Evaluation of the data on multiplicity fluctuations}
\label{sec:evaluation}

The shape of charged-particle multiplicity distributions changes with the acceptance $p<1$ of the detection region. The experimentally measured average multiplicity and variance of multiplicity distribution are expressed as:

\begin{equation}
\langle N \rangle=p\langle N_{p=1}\rangle,
\label{eq:mean}
\end{equation}
\begin{equation}
Var(N)=p^2Var(N_{p=1})+p(1-p)\langle N_{p=1}\rangle.
\label{eq:variance}
\end{equation}
Using Eqs.~(\ref{eq:mean}) and (\ref{eq:variance}) one can find that the variable:
\begin{equation}
\Omega=\frac{Var(N)}{\langle N\rangle^2}-\frac{1}{\langle N \rangle}  
\label{eq:omega1}
\end{equation}
does not depend on the detector acceptance $p$ \cite{GavinA}. This is essential for comparing different sets of experimental data. The so-called dynamic variance $\Omega$ was discussed in many applications. This function was connected with parameter $k$ from negative binomial distribution (NBD)~\footnote{$\Omega=-1/\langle N\rangle$ for $P\left(N\right)=\delta\left(N-\langle N\rangle\right)$, $\Omega=-1/K$ for binomial distribution, $\Omega=1/k$ for negative binomial distribution, and $\Omega=0$ for Poisson distribution.}, with non-extensivity parameter $q$, and with two-particle correlation function $\langle \nu_2\rangle$~\cite{Rybczynski:2004zi}:
\begin{equation}
\Omega=1/k=q-1=\langle \nu_2 \rangle.
\label{eq:Omega2}
\end{equation}

\begin{figure}[h]
 \begin{center}
\resizebox{0.45\textwidth}{!}{
\includegraphics{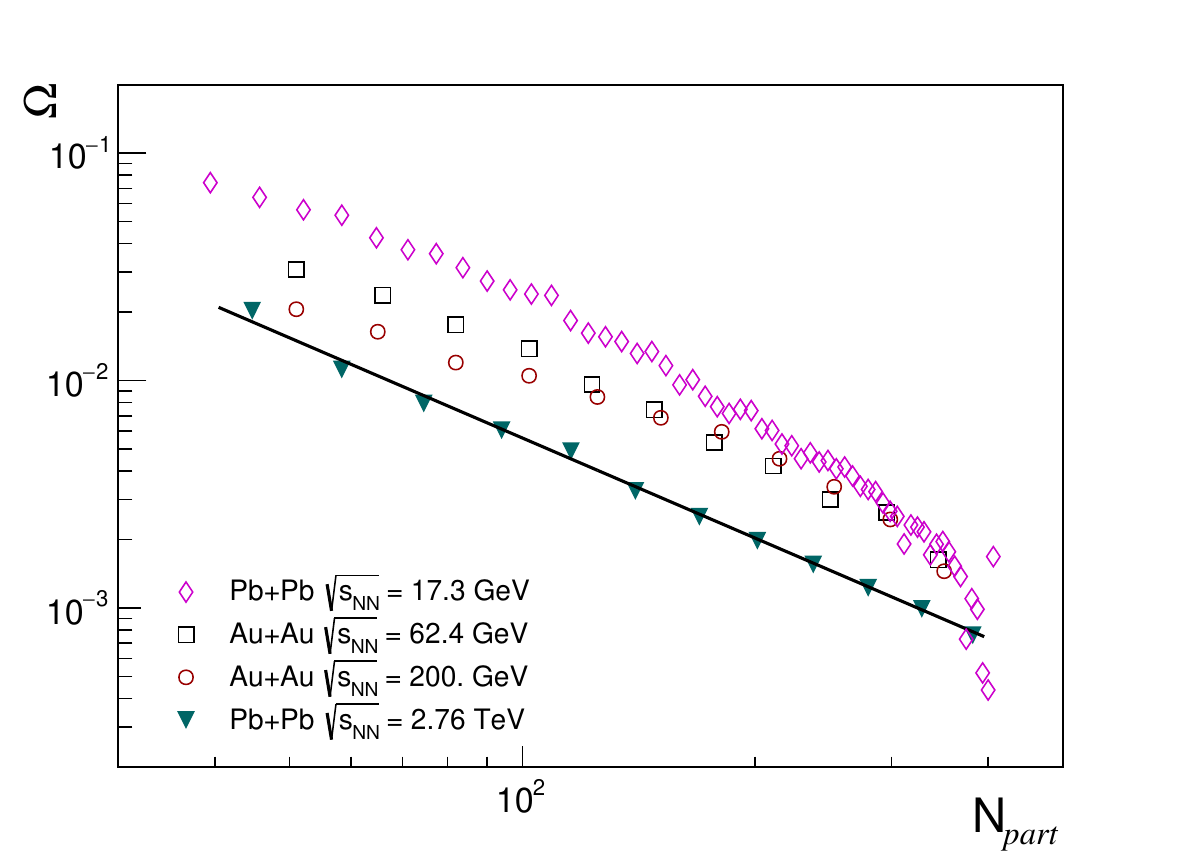}
}
\vspace*{0.3cm}
\caption{(Color online) Dependence of $\Omega=Var(N)/\langle N\rangle^2-1/\langle N\rangle$ on the number of participants $N_{part}$. The line show dependence: $\Omega =6 N^{-1.5}_{part}$. Based on data from: NA49~\cite{NA49-1} ($\sqrt{s_{NN}}=17.3$ GeV), PHENIX~\cite{PHENIX-1} ($\sqrt{s_{NN}}=62.4$ and $200$ GeV) and ALICE~\cite{ALICE-1} ($\sqrt{s_{NN}}=2760$ GeV) experiments.
} \label{Fig_1}
\end{center}
\end{figure}

In this paper we use experimental data on multiplicity distributions of charged particles produced in nuclear collisions obtained by the NA49~\cite{NA49-1} experiment at CERN SPS (Pb+Pb collisions at $\sqrt{s_{NN}}=17.3$ GeV), PHENIX~\cite{PHENIX-1} experiment at BNL RHIC (Au+Au collisions at $\sqrt{s_{NN}}=62.4$ and $200$ GeV) and ALICE~\cite{ALICE-1} experiment at CERN LHC (Pb+Pb collisions at $\sqrt{s_{NN}}=2760$ GeV). Using these data we first evaluate the values of $\Omega$. In Fig.~\ref{Fig_1} we show dependence of the dynamic variance $\Omega$ on the number of participants $N_{part}$. For the highest energy ($\sqrt{s_{NN}}=2.76$ TeV) $\Omega \sim N^{-1.5}_{part}$, but for lower energies we observe deviation from the simple power law dependence.

\section{Imprints of clustering}
\label{sec:imprints}

Let $N_{c}$ nucleons from a colliding nucleus interact collectively. The secondary particles produced in the interaction of $2N_c$ nucleons form the cluster. For $N_{part}$ nucleons participating in collision we have $ N_{S}=N_{part}/(2N_c)$ sources and each of these sources produce the average number of $\langle m\rangle$ particles with variance $Var(m)$. We can write:
\begin{equation}
Var \left(\sum^{N_S}_{i=1} m_i \right)= \sum^{N_S}_{i=1}Var(m_i)+\sum^{N_S}_{i=1}\sum^{N_S}_{j \neq i} Cov(m_i,m_j).
\label{eq:var_sum}
\end{equation}
The total number of secondary particles is $ N=\sum^{N_S}_{i=1} m_i $, and we rewrite: 
\begin{equation}
\Omega = \Omega'/N_S+\rho(\Omega'+1/\langle m\rangle)(1-1/N_S),
\label{eq:omega_sum}
\end{equation}
where the dynamic variance of the particles from one source $\Omega'$ is assumed to be the same for all sources and the correlation coefficient: 
\begin{equation}
\rho=\rho(m_i,m_j)=\frac{Cov(m_i,m_j)}{\sqrt{Var(m_i)Var(m_j)}}
\label{eq:rho}
\end{equation}
is assumed to be the same for all pairs of sources.

\begin{table*}
\caption{Parameters of the $\Omega N_{part}$ dependence on number of nucleons participating in collision, $N_{part}$ given by Eq.~(\ref{eq:fit}). The last column, $N_{c}$ contain the evaluated number of clusters shown in Fig.~\ref{Fig_Nc}.}
\label{tableforfitting}
\centering
\begin{tabular}{||c|c|c|c|c|c|c||}\hline
Exp. & Reaction &$\sqrt{s_{NN}}$~[GeV]&  a & b & $\gamma$& $N_c$\\ \hline
NA49 & $Pb+Pb$ &17.3 & 9.74 & 3.38 & 0.17 &9.6 \\ \hline
PHENIX &$Au+Au$ &62.4 & 3.242 & 0.569 &0.264& 3.8 \\ \hline
PHENIX & $Au+Au$  & 200 &1.6 &0.15&0.3&2.0 \\ \hline
ALICE & $Pb+Pb$ & 2760 &1.1 & 0.68 &0.387& 1.25\\ \hline
\end{tabular}
\end{table*} 

\begin{figure}[h]
\begin{center}
\resizebox{0.45\textwidth}{!}{
\includegraphics{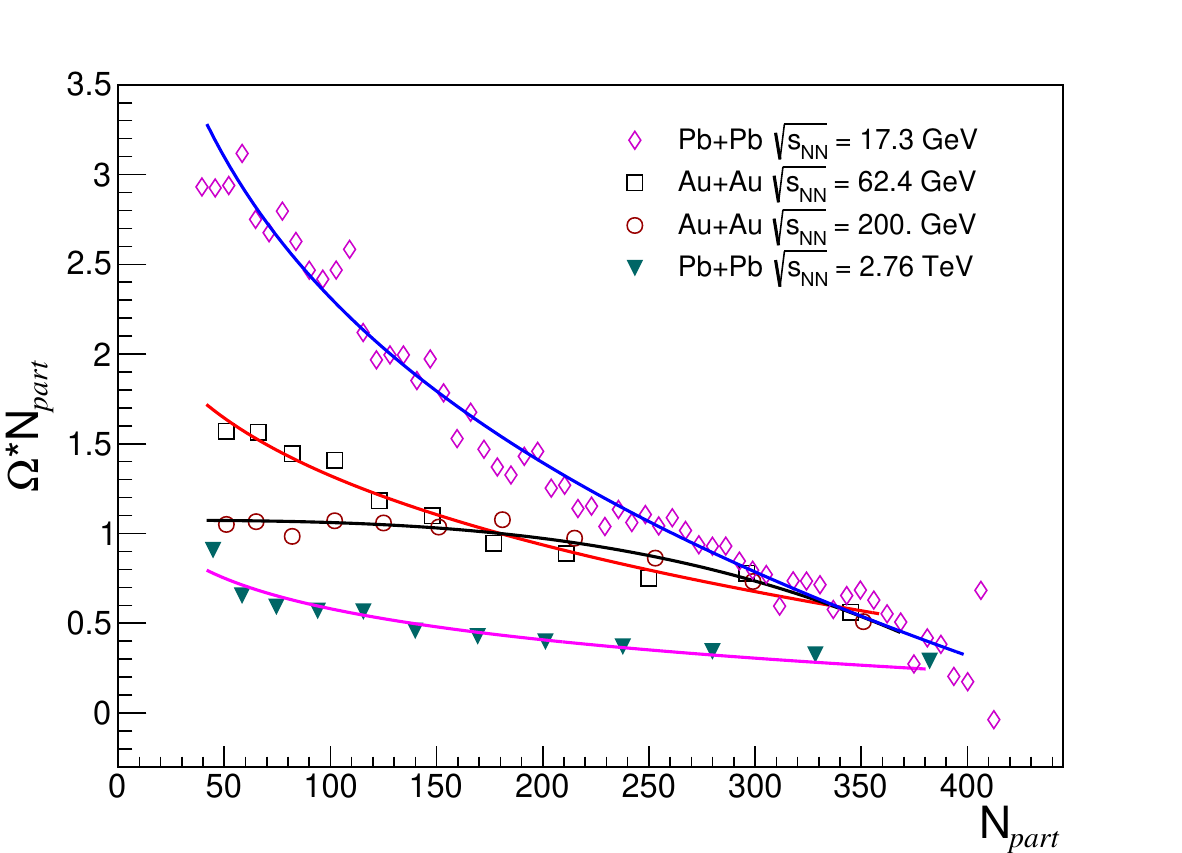}
}
\vspace*{0.3cm}
\caption{(Color online) $\Omega N_{part}$ as a function of the number of nucleons participating in collisions $N_{part}$. See text for details.
} \label{Fig_2}
\end{center}
\end{figure}

The experimental data can be fitted with the function: 
\begin{equation}
\Omega N_{part}= a-bN^{\gamma}_{part}
\label{eq:fit}
\end{equation}
with  parameters listed in the Table~\ref{tableforfitting}. In the simplest case we expect that $\Omega N_{part}=1-N_{part}/\left(2A\right)$ (see~\ref{sec:appendix} for more details). Deviation from the expected linear dependence on $N_{part}$ indicates that thermodynamic system is not in equilibrium with a heat bath. In the Fig.~\ref{Fig_2} we show dependence of the $\Omega N_{part}$ on the number of participants $N_{part}$ fitted by Eq.~(\ref{eq:fit}). \\

\begin{figure}[h]
\begin{center}
\resizebox{0.45\textwidth}{!}{
\includegraphics{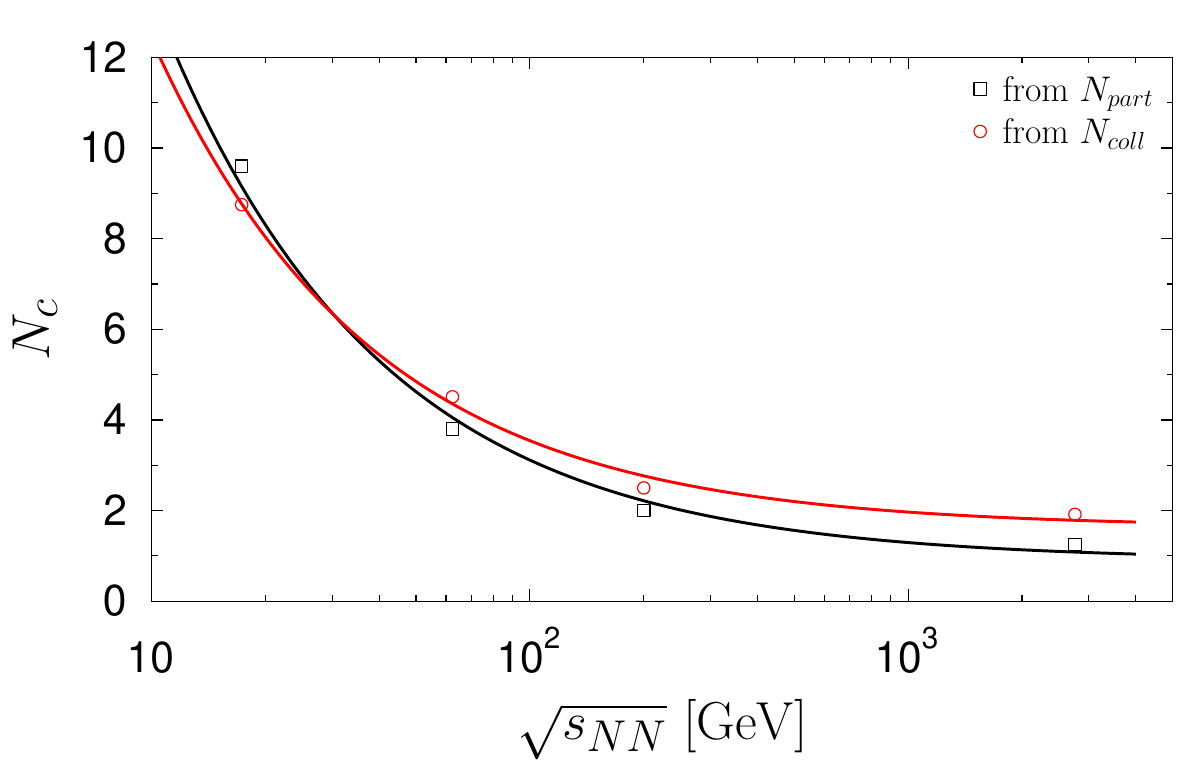}
}
\vspace*{0.3cm}
\caption{(Color online) Cluster size $N_{c}$ as a function of $\sqrt{s_{NN}}$ evaluated from $\Omega N_{coll}$ dependence on number of participants, $N_{part}$ and number of collisions, $N_{coll}$.
The energy dependencies follow: $N_c=0.9+70 (\sqrt{s_{NN}})^{-0.75}$ and $N_c=1.626+60.59 (\sqrt{s_{NN}})^{-0.75}$, respectively. See text for details.
} \label{Fig_Nc}
\end{center}
\end{figure}

In the nucleon-nucleon collisions, both mean multiplicity $\langle n_{pp}\rangle$ of charged particles and NBD parameter $k$ depend on the collision energy. The source from a cluster of the size $N_c$ produce on average $\langle n_{pp} \rangle N_c $ charged particles. Fluctuations of multiplicity exhibit self-similarity and the variance depend on mean multiplicity. Using the experimentaly obtained~\cite{Geich-Gimbel:1987zrl} energy dependence parametrizations of mean multiplicity, $\langle n_{pp}\left(\sqrt{s}\right)\rangle$ and NBD shape parameter, $k\left(\sqrt{s}\right)$ we get the values of NBD shape parameter as a function of the average charged particle multiplicity. Finally, following~\cite{Mryb} we obtain an NBD shape parameter $k$ as a function of the size of cluster $N_c$:
\begin{equation}
k^{-1} = \Omega' = -0.1+0.1 \sqrt{-1.356+0.503 \langle n_{pp}\rangle N_c}.
\label {eq:omega_i}
\end{equation}
The observed non-linear dependence of $\Omega N_{part}$ on the number of nucleons participating in collisions $N_{part}$ indicate that $\rho=\rho(N_{part})$ and only the first term in 
\begin{equation}
\Omega N_{part}=2 N_c \Omega'+ \rho\left(N_{part}\right)\left[\Omega'-\langle m\rangle^{-1}\right]\left[N_{part}-2N_c\right]
\end{equation}
is independent of $N_{part}$. Such regularity corresponds to description of experimental data given by Eq.~(\ref{eq:fit}). From equality $a=2N_c \Omega'$ we obtain the cluster size $N_c$. Energy dependence of the size of clusters $N_c$, shown in Fig.~\ref{Fig_Nc}, follows $N_c=0.9+70 (\sqrt{s_{NN}})^{-0.75}$.

The second term in Eq.~(\ref {eq:fit}) allows us to determine correlation coefficient:
\begin{equation}  
\rho=\frac{-bN^{\gamma}_{part}}{(N_{part}-2N_c)(\Omega'+ \langle m\rangle ^{-1})}.
\end{equation}

\section{Dependence on number of nucleon-nucleon collisions}
\label{sec:ncoll_dep}

\begin{table}
\caption{Parameters of the average number of binary nucleon-nucleon collisions, $\langle N_{coll}\rangle$ dependence on the number of nucleons participating in the collision, $\langle N_{part}\rangle$ given by Eq.~(\ref{eq:ncoll_npart}).}
\label{tableforfitting2}
\centering
\begin{tabular}{||c|c|c|c|c||}\hline
Reaction &$\sqrt{s_{NN}}$~[GeV]&  A & B & $\delta$ \\ \hline
$Pb+Pb$ &17.3 & 0.3 & 0.33 & 1.34  \\ \hline
$Au+Au$ &62.4 & 0.4 & 0.29 & 1.37 \\ \hline
$Au+Au$  & 200 & 0.5 & 0.25 & 1.42 \\ \hline
$Pb+Pb$ & 2760 & 0.6 & 0.22 & 1.5 \\ \hline
\end{tabular}
\end{table} 

\begin{table*}
\caption{Parameters of the $\Omega N_{coll}$ dependence on the average number of nucleons collisions, $\langle N_{coll}\rangle$ given by Eq.~(\ref{eq:fit_ncoll}). The last column, $N_{c}$ contain the evaluated number of clusters shown in Fig.~\ref{Fig_Nc}.}
\label{tableforfitting3}
\centering
\begin{tabular}{||c|c|c|c|c|c|c||}\hline
Exp. & Reaction &$\sqrt{s_{NN}}$~[GeV] &  $a_1$ & $b_1$ & $c_1$ & $N_c$\\ \hline
NA49 & $Pb+Pb$ & 17.3 & 4.19 & -0.00385 & $6.1\cdot 10^{-7}$ &8.75 \\ \hline
PHENIX &$Au+Au$ &62.4 & 2.15 & -0.00038 & $-4.5\cdot 10^{-7}$ & 4.51 \\ \hline
PHENIX & $Au+Au$  & 200 & 1.159 &0.00434 & $-3.95\cdot 10^{-6}$ & 2.5 \\ \hline
ALICE & $Pb+Pb$ & 2760 & 1.1 & 0.00034 & $-1.5\cdot 10^{-7}$ & 1.92\\ \hline
\end{tabular}
\end{table*} 

\begin{figure}[h]
\begin{center}
\resizebox{0.45\textwidth}{!}{
\includegraphics{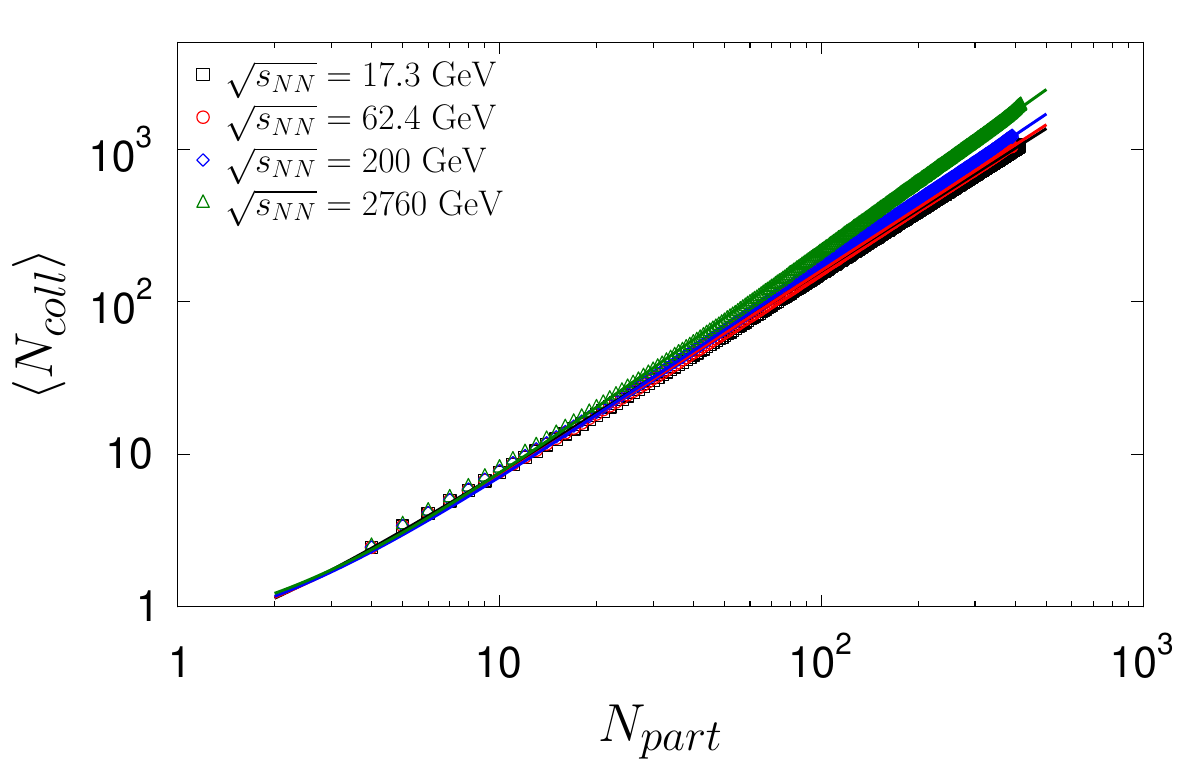}
}
\vspace*{0.3cm}
 \caption{(Color online) Average number of collisions, $\langle N_{coll}\rangle$ as a function of number of nucleons participating participating in the collisions evaluated from GLISSANDO~\cite{Broniowski:2007nz} simulation.
} \label{Fig_ncoll_npart}
\end{center}
\end{figure}

\begin{figure}[h]
\begin{center}
\resizebox{0.45\textwidth}{!}{
\includegraphics{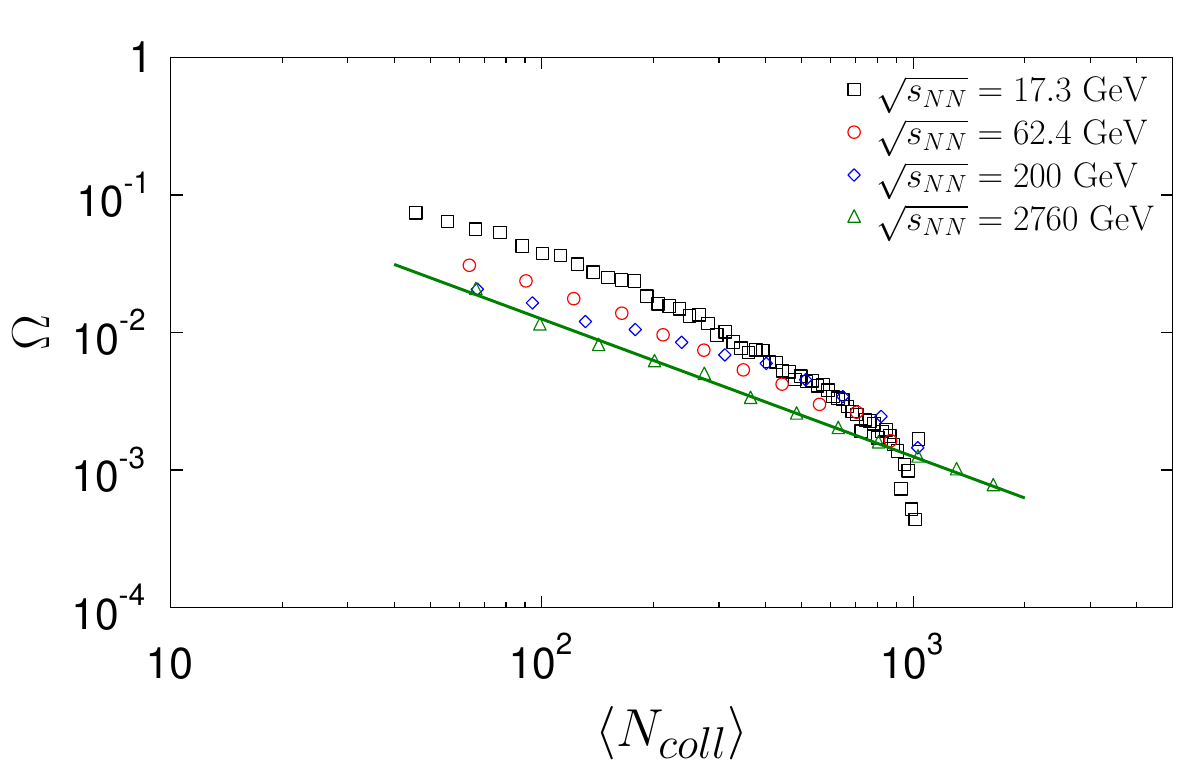}
}
\vspace*{0.3cm}
 \caption{(Color online) Dependence of  $\Omega=Var(N)/\langle N\rangle^2-1/\langle N\rangle$ on the average number of collisions, $\langle N_{coll}\rangle$. 
 The line show dependence: $\Omega =1.25 \langle N_{coll}\rangle^{-1.0}$.
} \label{Fig_Omega_ncoll}
\end{center}
\end{figure}

In this section we focus on studies of clustering dependence on the average number of binary nucleon-nucleon collisions, $\langle N_{coll}\rangle$ existing in the discussed interactions. We estimate $\langle N_{coll}\rangle$ as a function of number of nucleons participating in the collision using simulation done with Glauber-like Monte Carlo (GMC) generator, GLISSANDO~\cite{Broniowski:2007nz}. Fig.~\ref{Fig_ncoll_npart} shows the results fitted by the power-law dependence:
\begin{equation}
\langle N_{coll}\rangle=A+B\cdot N_{part}^{\delta}
\label{eq:ncoll_npart}
\end{equation}
with the parameters given in Table~\ref{tableforfitting2}. The GMC approach is very useful for the calculation of geometry related quantities like $N_{part}$ and $N_{coll}$. Within this approach $N_{part}$ scale with the volume of the interacting region. In a collision of two nuclei with the same number of nucleons the average number of collisions per participant nucleon scales with the length $L\sim N_{part}^{1/3}$ of the interaction volume along the beam direction. Thus, the number of collisions roughly follows $N_{coll}\sim N_{part}^{4/3}$ what does not depend on the size of colliding nuclei. The geometric nature of the GMC is obvious for collisions at low energies (small cross-section). The observed malformations in the estimate of $N_{part}$ and $N_{coll}$ indicate differences between optical and Monte Carlo approaches.

Having evaluated average numbers of collisions we use them to show in the Fig.~\ref{Fig_Omega_ncoll} the dynamic variance $\Omega$ dependence on $\langle N_{coll}\rangle$. For the highest energy ($\sqrt{s_{NN}}=2.76$ TeV) $\Omega \sim \langle N_{coll}\rangle^{-1}$, but for lower energies we observe deviation from the simple power-law dependence. 

\begin{figure}[h]
\begin{center}
\resizebox{0.45\textwidth}{!}{
\includegraphics{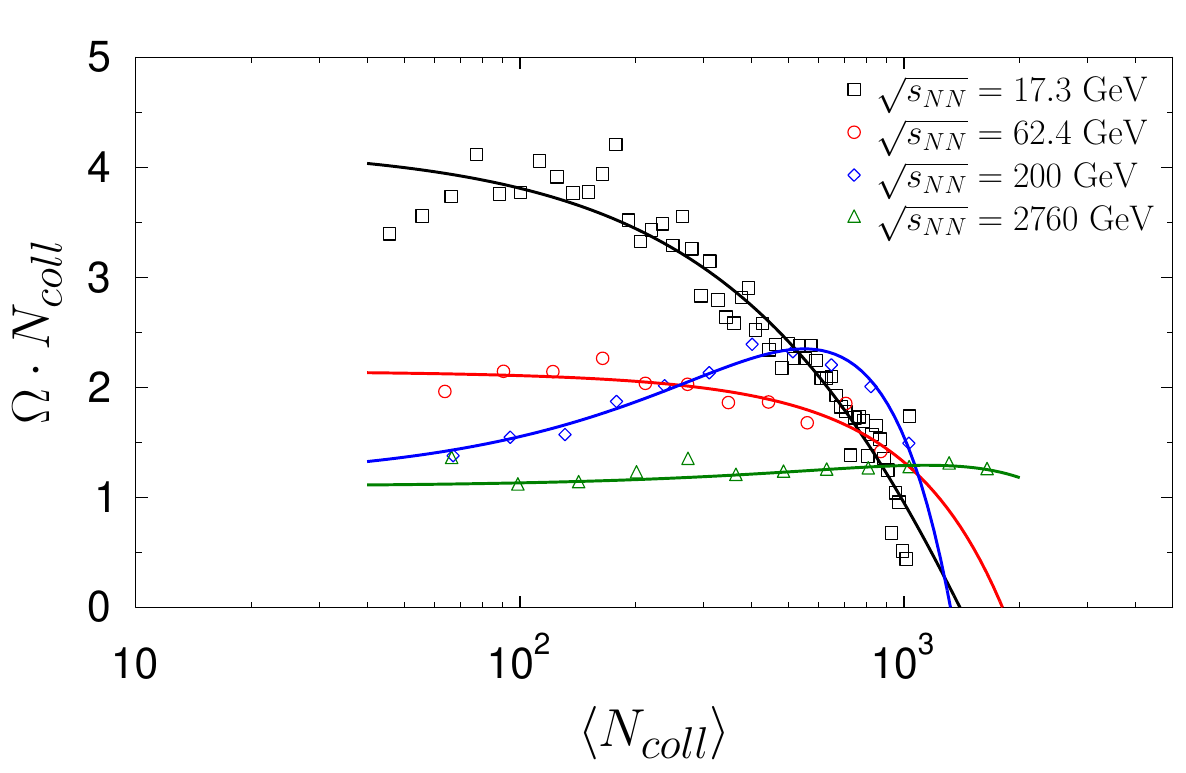}
}
\vspace*{0.3cm}
 \caption{(Color online) Dependence of  $\Omega N_{coll}$ on the average number of collisions, $\langle N_{coll}\rangle$. See text for details.
} \label{Fig_Omega_ncoll_vs_ncoll}
\end{center}
\end{figure}

Again, the experimental data can be fitted with the function: 
\begin{equation}
\Omega N_{coll}= a_1+b_1 \langle N_{coll}\rangle+c_1\langle N_{coll}\rangle^{2}
\label{eq:fit_ncoll}
\end{equation}
with parameters listed in the Table~\ref{tableforfitting3}. In Fig.~\ref{Fig_Omega_ncoll_vs_ncoll} we show dependence of the $\Omega N_{coll}$ on the average number of collisions $\langle N_{coll}\rangle$ fitted by Eq.~(\ref{eq:fit_ncoll}). The observed discrepancy between data and fit for the lowest collision energy, $\sqrt{s_{NN}}=17.3$~GeV can be caused by the non-monotonic dependence of the scaled variance of charged particle multiplicity distributions on the collision
centrality~\cite{NA49-1}.

Following procedure adopted previously in estimation of $N_{c}$, using values of $\langle N_{coll}\rangle$ we evaluate clusters sizes, $N_{c}=a_{1}/\Omega'$ for each discussed interaction. Energy dependence of the size of clusters $N_c$, shown in Fig.~\ref{Fig_Nc}, follows $N_c=1.626+60.59 (\sqrt{s_{NN}})^{-0.75}$. The size of the cluster does not depend significantly on the way it was evaluated.

\section{Summary}
\label{sec:summary}

In realistic heavy-ion collisions the extent light clusters, such as $N_{c}=4$ (tertahedron), $N_{c}=6$ (hexahedron), or $N_{c}=8$ (octahedron or square antiprism) can be formed. Clusters are the statistical correlation/association of $N_{c}$ nucleons appearing at the initial stage of the collision. Their energy has a large uncertainty and with overwhelming probability they form one source producing secondaries at the freeze-out stage. Experimental evidence for cluster formation coming from the statistical analysis of proton cumulants and higher-order moments (skewness and kurtosis) for collisions at the beam energy scan of RHIC has been proposed in~\cite{Shuryak:2018lgd}. Many models predict that baryon-rich matter will have the first order transition line, ending in a certain critical point~\cite{Stephanov1998,Stephanov1999,Koch:2001,Jeon2000,Jeon2003,Karsch:2005,Stokic:2008}. At the phase transition, the increase of fluctuations is expected~\cite{Stephanov1999,Karsch:2005,Baym:1989yf,Heiselberg:1991is,Blaettel:1992gu,Baym:1995cz}. However, the other sources of fluctuations coming from the initial stage cannot be forgotten, what at the high baryonic densities is poorly understood, so far.

In this work we have studied nucleonic clustering at the initial conditions corresponding to the baryon-rich heavy-ion collisions. More specifically, we have observed that both the clustering rate and the properties of the resulting clusters are very sensitive to the fluctuations of secondary produced hadrons, and suggest that detailed studies of such behaviour will allow ultimately tell us whether the QCD critical point exist or not. Our analysis suggest that nucleon clusters are produced in experiment. The size of clusters change from $N_{c}=9$ to $N_{c}=1$ when the interaction energy ranges from tens of GeV up to a few of TeV.


\begin{acknowledgements}
This research was supported by the Polish National Science Centre (NCN) Grant No. 2020/39/O/ST2/00277. In preparation of this work we used the resources of the Center for Computation and Computational Modeling of the Faculty of Exact and Natural Sciences of the Jan Kochanowski University of Kielce. 
\end{acknowledgements}

\appendix

\section{Finiteness of source and reservoir}
\label{sec:appendix}

The statistical model of hadronization (SMH) is a very useful tool for the description of average particle multiplicities in reactions induced by relativistic ions~\cite{Cleymans:1992,Yengranddon1997,Becattini:1997} as well as in elementary particle reactions~\cite{Becattini:1997, Becattini:1997:23, Becattini:2001:24} Within SMH it is possible to estimate multiplicity fluctuations due to the fact that the status of the hadronizing sources is well defined.\\

Some time ago electric charge and charged-particle multiplicity fluctuations have been proposed as a tool to distinguish between hadron gas and QGP~\cite{Jeon2000,Asakawa:2000}. The conservation constraints on fluctuations observed in thermal ensembles of relativistic ion collisions were first discussed in~\cite{Stephanov1999}. Then it was shown~\cite{Begun2005,BegunV2004} that in the canonical ensemble (CE) with exact conservation of charge, the scaled variance of multiplicity distribution of any particle type does not converge to the corresponding grand-canonical (GCE) value even in the thermodynamic limit, unlike the mean~\cite{Cleymansnov1997, Kernanen2002}. \\

Let us divide a CE with a large volume into a cluster, which is GCE with the remaining part of the system being treated as a reservoir~\cite{Becattini:32}. Multiplicity distribution in a cluster is yielded by a negative binomial distribution. The variance of a multiplicity distribution, $Var(N)$, hinges on the average multiplicity $\langle N\rangle$ in the system. When $\langle N\rangle$ increases with energy, $Var(N)$ also changes.\\

Temperature fluctuations present in a finite system being in contact with finite thermal bath was considered in Ref.~\cite{HBProsper1:1993} when analyzed combinatorial problem of counting number of microstates of the combined system of ensemble and bath. For a system which is formed in collision of $N_{part}$ nucleons being in contact with $2A-N_{part}$ spectators we have
\begin{equation}
q-1=\frac{Var(1/T)}{\langle 1/T\rangle ^2} =\frac{1}{N_{part}}(1-\frac{N_{part}}{2A}).
\label{eq:q_Prosper}
\end{equation}
For a finite size system being in contact with a heat bath, using Lindhard’s approach \cite{Linhard:33}, we have:
\begin{equation}
Var\left(U\right)+C_V^2 Var(T) =\langle T \rangle ^2C_V.
\label{eq:varU}
\end{equation}
Relation (\ref{eq:varU}) is supposed to be valid all the way from the canonical ensemble, where $ Var(T ) = 0$ and $Var(U) =\langle T \rangle^2 C_V$, to the microcanonical ensemble, for which $Var(T ) =\langle T \rangle^2/C_V$ and $Var(U) = 0$. Eq.~(\ref{eq:varU}) expresses the complementarity between all the temperature and energy and the canonical and microcanonical description of the system~\cite{Uffink:34,Campisi:35}.
For the intermediate case it can be assumed that:
\begin{equation}
Var\left(U\right)=\langle T \rangle^2C_V f(N_ {part})
\label{eq:varU_2}
\end{equation}
and we obtain
\begin{equation}
q-1 \sim \frac{1}{N_{part}}(1-f(N_{part}) ).
\label{eq:q1}
\end{equation}

In the simplest case, $f(N_{part})=N_{part}/(2A)$ we have Eq.~(\ref{eq:q_Prosper}) which can describe multiplicity fluctuations of secondaries produced in Pb+Pb collisions at $\sqrt{s_{NN}}=17.3 $ GeV~\cite{Wilk:2009nn}. Usually, a thermodynamic system is not in equilibrium with a heat bath, and we can expect that
\begin{equation}
f\left(N_{part}\right)=bN^\gamma_{part}
\label{eq:f}
\end{equation}
corresponds with Eq.~(\ref{eq:fit}).


\end{document}